\documentclass[aps,twocolumn,showpacs]{revtex4} 

\newcommand{\stac}[2]{\stackrel{\scriptscriptstyle {#1}}{#2}}

\begin{document}
%\draft

%<<<<<<<<<<<<< TITLE >>>>>>>>>>>>>>>%
\title{Gravity is controlled by cosmological constant}

%<<<<<<<<<<<<< AUTHOR >>>>>>>>>>>>>>>%
\author{Yukinori Iwashita$^{(1)}$, Tetsuya Shiromizu $^{(1,2,3)}$, 
Keitaro Takahashi$^{(2)}$ and Shunsuke Fujii$^{(1)}$}

%<<<<<<<<<<<<< ADDRESS >>>>>>>>>>>>>>>%
\affiliation{$^{(1)}$Department of Physics, Tokyo Institute of Technology, 
Tokyo 152-8551, Japan}

\affiliation{$^{(2)}$Department of Physics, The University of Tokyo,  Tokyo 
113-0033, Japan}

\affiliation{$^{(3)}$Advanced Research Institute for Science and Engineering, 
Waseda University, Tokyo 169-8555, Japan}

%<<<<<<<<<<<<< DATE >>>>>>>>>>>>>>>%
\date{\today}

%======================================%
%<<<<<<<<<<<<< ABSTRACT >>>>>>>>>>>>>>>%
%======================================%
\begin{abstract}
We discuss a Randall-Sundrum-type two D-braneworld model in which D-branes possess different values
of the tensions from those of the charges, and derive an effective gravitational equation 
on the branes. As a consequence, the Einstein-Maxwell theory is realized together with
the non-zero cosmological constant. Here an interesting point is that the effective gravitational
constant is proportional to the cosmological constant. If the distance between two D-branes
is appropriately tuned, the cosmological constant can have a consistent value with
the current observations. From this result we see that, in our model, the presence of
the cosmological constant is naturally explained by the presence of the effective gravitational
coupling of the Maxwell field on the D-brane. 
\end{abstract}


\maketitle
%\vskip1cm

%======================================%
%<<<<<<<<<<<< SECTION I  >>>>>>>>>>>>>>%
%======================================%
%\baselineskip25pt
\label{sec:intro}
\section{Introduction}

Recent progress of the superstring theory provides us a new picture of the universe. 
This is so called braneworld: our universe is described as a thin domain wall
in higher dimensional spacetime (See Ref \cite{Review} for recent reviews). 
While the braneworld model is strongly motivated by the superstring theory,
most of the previously studied models remain rather unrealistic in the context
of D-brane in the superstring theory, although some features of D-brane have
been taken into account in the studies of probe D-brane cosmology like mirage 
cosmology \cite{Mirage} (See Refs. \cite{DBW1, DBW2, DBW3} for other related issues.). 

Recently, a more realistic inflation model has been proposed in $D \bar D$ system
with a flux compactification \cite{Dbrane}. In this model, the inflaton corresponds to the 
radion, which represents the distance between the branes. Our universe may be considered
to a D3-brane separated from the $D \bar D$ system. Anyway, the description of
our universe is slightly rough because the effective gravitational equation on the branes
was not discussed carefully.

In such situation one of the authors tackled this issue with his collaborators
\cite{SKOT, OSKH, SHT}. The purpose there was of course the effective gravitational theory
on the D-brane. The bulk spacetime and brane were described by the bosonic part of the 
ten-dimensional IIB supergravity theory compactified on $S^5$ and the Born-Infeld plus
Chern-Simons action, respectively. The brane tension was set equal to the brane charge
and $Z_2$ symmetry was assumed. This setup can be considered as a type IIB supergravity extension
of the Randall-Sundrum model \cite{RSI, RSII}. 

Intuitively we expected that the effective gravitational equation
on the brane would be the Einstein-Maxwell theory because the Born-Infeld action includes
a $U(1)$ field. However, this was not the case. We obtained a conclusion that the Maxwell field
localized on the brane does not contribute as a source for the gravity on the brane.
Then we suspected that the BPS condition (the brane charge $=$ brane tension) would be severe,
and discussed a non-BPS case with a single non-BPS D-brane \cite{SKT}.
As a result, it turned out that the gauge field could be source for the gravity on the brane.
However, there was also an anomalous term in the effective energy-momentum tensor,
that is, a trace term exists as
$g_{\mu\nu} F_{\alpha\beta} F^{\alpha\beta} \in T_{\mu\nu}^{\rm eff}$. 
Here we should stress that this non-BPS case would be the only possible one to realize
a coupling of the Maxwell field to the gravity. Indeed, we obtained essentially
the negative conclusion for a $Z_2$ asymmetric case \cite{TS}. 

In this paper, we reexamine the non-BPS D-brane model, not with a single brane
but two branes. When one uses the gradient expansion method \cite{GE} to solve the bulk spacetime,
the single brane case has an ambiguity which corresponds to an ``integration constant'',
which is actually a function of the brane coordinates. Physically, this may be regarded
as a holographic dark radiation \cite{GE,SI}, although this is an non-trivial issue.
Anyway, there remains an unknown term in the effective Einstein equation unless
the integration constant is fixed by another boundary condition in the bulk spacetime.
We guess that this is the reason why the anomalous term appears in the
energy-momentum tensor of the effective theory for the single non-BPS D-brane model \cite{SKT}. 
In order to get a reliable effective theory, we have to compute the integration constant
carefully by putting another D-brane and imposing a boundary condition there.
Then the integration constant will be completely determined and there remains no ambiguity.
Hence we can expect that we will be able to obtain a definite answer for the gravitational theory 
on the brane. 

The rest of this paper is organized as follows. In Sec. II we describe our 
toy model which is a simplification of the type IIB supergravity theory and give
the basic equations in Sec. III. In Sec. IV, we solve the bulk spacetime
with the boundary conditions, which are given by the junction conditions on the two branes. 
Then we derive an effective gravitational equation on the branes. 
Finally we will give the summary and discussions in Sec. V.

%======================================%
%<<<<<<<<<<<< SECTION II  >>>>>>>>>>>>>%
%======================================%
%\baselineskip25pt
\section{Model}
\label{sec:model}

We consider a Randall-Sundrum type model inspired by the type IIB supergravity compactified on 
$S^5$. The brane is described by the Born-Infeld and Chern-Simons actions. So we begin with 
the following total action 
%===========<Equation>============%
%
\begin{eqnarray}
S & = & \frac{1}{2\kappa^2} \int d^5x {\sqrt {-{\cal G}}}\biggl[{}^{(5)}R-2\Lambda 
-\frac{1}{2}|H|^2 
\nonumber \\ 
& & -\frac{1}{2}(\nabla \chi)^2-\frac{1}{2}|\tilde F|^2-\frac{1}{2}|\tilde 
G|^2 \biggr] \nonumber \\ 
& & +S_{\rm brane}^{(+)}+S_{\rm CS}^{(+)}+S_{\rm brane}^{(-)}+S_{\rm CS}^{(-)} , 
\label{action} 
\end{eqnarray} 
%
%=================================%
where $H_{MNK}=\frac{1}{2}\partial_{[M}B_{NK]}$, 
$F_{MNK}=\frac{1}{2}\partial_{[M}C_{NK]}$, 
$G_{K_1 K_2 K_3 K_4 K_5}=\frac{1}{4!}\partial_{[K_1}D_{K_2 K_3 K_4 K_5]}$, 
$\tilde F = F + \chi H$ and $\tilde G=G+C \wedge H$. $M,N,K=0,1,2,3,4$. 
$B_{MN}$ and $C_{MN}$ are 2-form fields, and $D_{K_1 K_2 K_3 K_4}$ is
a 4-form field. $\chi$ is a scalar field. ${\cal G}_{MN}$ is the metric 
of five dimensional spacetime. $A_{[MN]}:=A_{MN}-A_{NM}$. $\Lambda$ is the 
bulk negative cosmological constant. 

$S_{\rm brane}^{(\pm)}$ is given by the Born-Infeld action
%===========<Equation>============%
%
\begin{eqnarray}
S_{\rm brane}^{(\pm)}=-\beta_{(\pm)} \int d^4x {\sqrt {-{\rm det}(g_{(\pm)}+{\cal F}^{(\pm)})}}, 
\end{eqnarray}
%
%=================================%
where $g_{(\pm) \mu\nu}$ are the induced metric on the $D_{\pm}$-brane and 
%===========<Equation>============%
%
\begin{eqnarray}
{\cal F}_{\mu\nu}^{(\pm)}=B_{\mu\nu}^{(\pm)}+(|\beta_{(\pm)}|)^{-1/2}F_{\mu\nu}^{(\pm)}.
\end{eqnarray}
%
%=================================%
$F_{\mu\nu}$ is the $U(1)$ gauge field on the brane. Here $\mu,\nu=0,1,2,3$ and 
$\beta_{(\pm)}$ are $D_{\pm}$-brane tension. 

$S_{\rm CS}^{(\pm)}$ is the Chern-Simons action 
%===========<Equation>============%
%
\begin{eqnarray}
S_{\rm CS}^{(\pm)} & = & -\gamma_{(\pm)} \int d^4x {\sqrt {-g_{(\pm)}}} 
\epsilon^{\mu\nu\rho\sigma}\biggl[ \frac{1}{4}{\cal 
F}_{\mu\nu}^{(\pm)}C_{\rho\sigma}^{(\pm)} \nonumber \\
& & +\frac{\chi}{8}{\cal F}_{\mu\nu}^{(\pm)}{\cal F}_{\rho\sigma}^{(\pm)}
+\frac{1}{24}D_{\mu\nu\rho\sigma}^{(\pm)} \biggr], 
\end{eqnarray}
%
%=================================%
where $\gamma_{(\pm)}$ are the brane charges. 
Here the brane charges are not equal to the brane tensions in general. Therefore, our model 
can contain non-BPS state of D-branes and we are interested in such branes. 

In this paper we deal with the toy model above. The difference from the original model 
based on the type IIB supergravity theory \cite{SKOT} is the absence of the scalar fields
related to the dilaton and the $S^5$ compactification. Instead, we introduced the bulk
cosmological constant $\Lambda$ in order to realise the similar features to the 
original model. Also
it should be noted that the scalar fields are not essential in considering the coupling
between the Maxwell field and the gravity \cite{SKOT}. In the original model, the brane tension
was set equal to the brane charge and then a flat brane was contained as a solution. 
To realize a flat brane under BPS condition $\gamma_{(\pm)}=\beta_{(\pm)}$ in our toy model,
we assume the following relation between the bulk cosmological constant and the brane charge:
%===========<Equation>============%
%
\begin{eqnarray}
2\Lambda = -\frac{5}{6}\kappa^4 \gamma_{(+)}^2. 
\end{eqnarray}
%
%=================================%

%=======================================%
%<<<<<<<<<<<< SECTION III  >>>>>>>>>>>>>%
%=======================================%

\section{Basic equations}

In this section we write down the basic equations and boundary conditions. 
Let us perform (1+4)-decomposition 
%===========<Equation>============%
%
\begin{eqnarray}
ds^2={\cal G}_{MN}dx^{M}dx^{N}=dy^2+g_{\mu\nu}(y,x) dx^\mu dx^\nu.
\end{eqnarray}
%
%=================================%
where $D_+$-brane and $D_-$-brane are supposed to be located at $y=\phi_+ (x)$ and
$y=\phi_- (x)$ \cite{moduli}. 
 
The spacelike ``evolutional" equations to the $y$-direction are 
%===========<Equation>============%
%
\begin{eqnarray}
\partial_y K 
& = & {}^{(4)} R-\kappa^2 \biggl( {}^{(5)}T^\mu_\mu -\frac{4}{3}{}^{(5)}T^M_M \biggr) -K^2, 
\label{evoK}
\end{eqnarray}
%
%=================================%
%===========<Equation>============%
%
\begin{eqnarray}
\partial_y \tilde K^\mu_\nu & = &  {}^{(4)}\tilde R^\mu_\nu 
-\kappa^2\biggl({}^{(5)}T^\mu_\nu 
-\frac{1}{4} 
\delta^\mu_\nu {}^{(5)}T^\alpha_\alpha \biggr)-K \tilde K^\mu_\nu, 
\label{traceless} 
\end{eqnarray}
%
%=================================%
%===========<Equation>============%
%
\begin{eqnarray} 
\partial_y^2 \chi +D^2 \chi +K\partial_y \chi-\frac{1}{2}H_{y\alpha\beta}\tilde 
F^{y\alpha\beta}=0, 
\end{eqnarray} 
%
%=================================%
%===========<Equation>============%
%
\begin{eqnarray}
\partial_y X^{y\mu\nu}+KX^{y\mu\nu}+D_\alpha H^{\alpha\mu\nu} +\frac{1}{2}F_{y\alpha\beta}\tilde G^{y\alpha\beta\mu\nu}=0, 
\label{evoH} 
\end{eqnarray} 
%
%=================================%
%===========<Equation>============%
%
\begin{eqnarray}
\partial_y \tilde F^{y\mu\nu}+K \tilde F^{y\mu\nu}
+D_\alpha \tilde F^{\alpha\mu\nu}-\frac{1}{2}H_{y\alpha\beta}
\tilde G^{y\alpha\beta\mu\nu}=0,
\label{evoF}
\end{eqnarray}
%
%=================================%
%===========<Equation>============%
%
\begin{eqnarray}
\partial_y \tilde G_{y \alpha_1 \alpha_2 \alpha_3 \alpha_4}
= K\tilde  G_{y \alpha_1 \alpha_2 \alpha_3 \alpha_4},
\end{eqnarray}
%
%=================================%
where $X^{y\mu\nu}:=H^{y\mu\nu}+\chi \tilde F^{y\mu\nu}$ and the 
energy-momentum tensor is 
%===========<Equation>============%
%
\begin{eqnarray}
&& \kappa^2\;{}^{(5)\!}T_{MN} =  \frac{1}{2}\biggl[ \nabla_M \chi \nabla_N \chi
-\frac{1}{2}g_{MN} (\nabla \chi)^2 \biggr]
\nonumber \\
& & ~~~~~~~~~~
+\frac{1}{4}\biggl[H_{MKL}H_N^{~KL}-g_{MN}|H|^2 \biggr] 
\nonumber \\
& & ~~~~~~~~~~
 +\frac{1}{4}\biggl[\tilde F_{MKL}\tilde
F_N^{~KL}-g_{MN}|\tilde F|^2
\biggr]
\nonumber \\
& & ~~~~~~~~~~
 +\frac{1}{96}\tilde G_{MK_1 K_2 K_3 K_4} \tilde G_{N}^{~~K_1
K_2 K_3 K_4}-\Lambda g_{MN}.
\nonumber \\
& & 
\end{eqnarray}
%
%=================================%
$K_{\mu\nu}$ is the extrinsic curvature, $K_{\mu\nu}=\frac{1}{2} \partial_y g_{\mu\nu}$. 
$\tilde K^\mu_\nu$ and ${}^{(4)}\tilde R^\mu_\nu$ are the traceless parts 
of $K^\mu_\nu$ and ${}^{(4)}R^\mu_\nu$, respectively. 
Here $D_\mu$ is the covariant derivative with respect to $g_{\mu\nu}$.

The constraints on $y={\rm const.}$ hypersurfaces are 
%===========<Equation>============%
%
\begin{eqnarray}
& & -\frac{1}{2}\biggl[{}^{(4)}R-\frac{3}{4}K^2+\tilde K^\mu_\nu \tilde K^\nu_\mu \biggr]
=\kappa^2\:{}^{(5)\!}T_{yy}, 
\label{conK}
\end{eqnarray}
%
%=================================%
%===========<Equation>============%
%
\begin{eqnarray}
D_\nu K^\nu_\mu-D_\mu K = \kappa^2\:{}^{(5)\!}T_{\mu y},
\end{eqnarray}
%
%=================================%
%===========<Equation>============%
%
\begin{eqnarray}
D_\alpha X^{y\alpha\mu}+\frac{1}{6} F_{\alpha_1 \alpha_2 \alpha_3} 
\tilde G^{y \alpha_1 \alpha_2 \alpha_3 \mu}= 0, \label{con1}
\end{eqnarray}
%
%=================================%
%===========<Equation>============%
%
\begin{eqnarray}
D_\alpha  \tilde F^{y\alpha\mu}-\frac{1}{6} H_{\alpha_1 \alpha_2 \alpha_3}
\tilde G^{y \alpha_1 \alpha_2 \alpha_3 \mu}=0, \label{con2}
\end{eqnarray}
%
%=================================%
%===========<Equation>============%
%
\begin{eqnarray}
D^\alpha \tilde G_{y \alpha \mu_1 \mu_2 \mu_3}=0.
\end{eqnarray}
%
%=================================%

Under $Z_2$-symmetry, the junction conditions at the brane located $y=\phi_\pm (x)$ are 
%===========<Equation>============%
%
\begin{eqnarray}
& &  K^{\mu}_\nu(\phi_\pm,x)  \nonumber \\
& & ~~= \mp {\sqrt {1+g^{\rho \sigma}_{(\pm)} \partial_\rho \phi_\pm \partial_\sigma \phi_\pm }}
\Bigl(  \frac{\kappa^2}{6} \beta_{(\pm)} \delta^\mu_\nu  + \frac{\kappa^2}{2} \beta_{(\pm)}
T^{\mu (\pm)}_{\nu}\Bigr) \nonumber \\
& & ~~~~
+\Bigl(D^\mu D_\nu \phi_\pm \pm 
\frac{\kappa^2}{6}\beta_{(\pm)} D^\mu \phi_\pm D_\nu \phi_\pm \Bigr) 
\nonumber \\
& & ~~~~+O(T_{\mu\nu}^2) \label{omit} \\ 
& & H_{y\mu\nu}(\phi_\pm,x)=\pm \kappa^2 \beta_{(\pm)}  {\cal F}_{\mu\nu}^{(\pm)}, \\ 
& & \tilde F_{y\mu\nu}(\phi_\pm,x)
=\pm \frac{\kappa^2}{2}\gamma_{(\pm)}  \epsilon_{\mu\nu\alpha\beta}{\cal F}^{(\pm)\alpha\beta}, \\ 
& & \tilde G_{y\mu\nu\alpha\beta}(\phi_\pm,x) 
=\pm \kappa^2 \gamma_{(\pm)}  \epsilon_{\mu\nu\alpha\beta},\\ 
& & \partial_y \chi (\phi_\pm,x) 
= \pm \frac{\kappa^2}{8}\gamma_{(\pm)}  \epsilon^{\mu\nu\alpha\beta}{\cal F}^{(\pm)}_{\mu\nu}{\cal 
F}_{\alpha\beta}^{(\pm)}.
\end{eqnarray}
%
%=================================%
In the above 
%===========<Equation>============%
%
\begin{eqnarray}
T^{(\pm)\mu}_{~~~~~\nu}={\cal F}^{(\pm)\mu\alpha}{\cal F}^{(\pm)}_{\nu \alpha} -\frac{1}{4}\delta^\mu_\nu 
{\cal F}_{\alpha\beta}^{(\pm)} {\cal F}^{(\pm) \alpha\beta}
\end{eqnarray}
%
%=================================%
and we discarded the higher order terms which will be negligible under the assumption of Eq. (\ref{eq:hierarchy}). 

From the junction condition for $\chi$, we can omit the contribution of $\chi$ to the gravitational 
equation on the brane in the approximations which we will employ. Moreover, we omit the quadratic 
term of the energy-momentum tensor in Eq. (\ref{omit}). 

For simplicity, we impose $\tilde F_{\mu\nu\alpha}=0$ and $H_{\mu\nu\alpha}=0$. 
Also we assume the deviation from BPS state is small ($|\gamma| \gg |\gamma - \beta|$).

%======================================%
%<<<<<<<<<<<< SECTION IV  >>>>>>>>>>>>>%
%======================================%
\section{Gradient expansion and effective equation}

The derivation of the gravitational equation on D-branes is our end here. The 
geometrical projection method developed in Ref. \cite{SMS} is one of the powerful tools
to see the effective equation. This is because we can derive an effective equation 
without solving the full spacetime. The equation contains a contribution from the 
bulk spacetime in the form of the projected five dimensional Weyl tensor. If the 
cosmological constant is the only bulk matter, the contribution will be negligible 
at low-energy scales. However, we have a caution for the case with bulk fields.
Indeed, we cannot omit the contribution from the Weyl tensor even at low-enegy 
scales and then we must solve the bulk spacetime to evaluate the contribution from
the Weyl tensor. 

To obtain the effective theory on the brane, therefore, 
we first solve the bulk spacetime. We will use the long wave approximation \cite{GE}. 
The small parameter is the ratio of the bulk curvature scale $\ell$ to the brane
intrinsic curvature scale $L$ due to the ordinary matter origin:
%===========<Equation>============%
%
\begin{eqnarray}
\epsilon= \frac{\ell^2}{L^2} \ll 1.
\end{eqnarray}
%
%=================================%
In addition we assume the following ordering 
%===========<Equation>============%
%
\begin{eqnarray}
|\gamma| \gg
|\gamma-\beta| > |\beta T_{\mu\nu}^{(\pm)}| > |\beta ({\cal D}\phi)^2 |, 
|\beta \ell {\cal D}^2 \phi |.
\label{eq:hierarchy}
\end{eqnarray}
%
%=================================%

The bulk metric is written as 
%===========<Equation>============%
%
\begin{eqnarray}
ds^2=dy^2+g_{\mu\nu}(y,x)dx^\mu dx^\nu, 
\end{eqnarray}
%
%=================================%
In this coordinate, the brane is supposed to be located at $y=\phi_\pm(x)$. Therefore 
the induced metric $g_{(\pm)\mu\nu}$ becomes 
%===========<Equation>============%
%
\begin{eqnarray}
g_{(\pm) \mu\nu}=g_{\mu\nu}(\phi_\pm,x)+\partial_\mu \phi_\pm (x) \partial_\nu \phi_\pm (x).
\end{eqnarray}
%
%=================================%

In the gradient expansion, the metric and the extrinsic curvature are expanded as 
%===========<Equation>============%
%
\begin{eqnarray}
g_{\mu\nu}(y,x) = a^2(y) \Bigl[h_{\mu\nu}(x)+\stac{(1)}{g}_{\mu\nu}(y,x)+\cdots \Bigr]
\end{eqnarray}
%
%=================================%
and 
%===========<Equation>============%
%
\begin{eqnarray}
K^\mu_\nu = \stac{(0)}{K^\mu_\nu}+ \stac{(1)}{K^\mu_\nu}+\cdots, 
\end{eqnarray}
%
%=================================%
where we set $\stac{(1)}{g}_{\mu\nu}(\phi_+,x)=0$. 

The strategy for obtaining the effective gravitational equation on the brane is as follows. 
We first solve the bulk spacetime using gradient expansion and compute the extrinsic curvatures. 
We can expect $K_{\mu\nu}(y,x) \sim {}^{(4)}R_{\mu\nu} + \cdots $. 
Applying the junction condition ($K_{\mu\nu}(y_\pm, x) \sim T_{\mu\nu}$) at the branes, 
we can obtain the effective equation on the branes. Intuitively, we expect that
the four dimensional Einstein equation can be realized in the first order. However,
this is not the case of our model. Therefore we need to compute the second order perturbations
to see if the conventional gravitational theory is reproduced.

%--------------------------------------%
%<<<<<<<<<< Subsection A  >>>>>>>>>>>>>%
%--------------------------------------%
\subsection{Background}

For the background spacetime, the evolutional equations are 
%===========<Equation>============%
%
\begin{eqnarray}
 \partial_y \stac{(0)}{K} = -\kappa^2 \Biggl({}^{(5)}T^\mu_\mu 
-\frac{4}{3}{}^{(5)}T^M_M \Biggr)^{(0)} -\stac{(0)}{K^2},
\end{eqnarray}
%
%=================================%
%===========<Equation>============%
%
\begin{eqnarray}
\partial_y \stac{(0)}{\tilde K^\mu_\nu}=-\stac{(0)}{K}\stac{(0)}{\tilde K^\mu_\nu}
\end{eqnarray}
%
%=================================%
and
%===========<Equation>============%
%
\begin{eqnarray}
\partial_y \stac{(0)}{\tilde G}_{y \alpha_1 \alpha_2 \alpha_3 \alpha_4} =  \stac{(0)}{K}
\stac{(0)}{\tilde G}_{y \alpha_1 \alpha_2 \alpha_3 \alpha_4}.
\end{eqnarray}
%
%=================================%
The constraint equations are 
%===========<Equation>============%
%
\begin{eqnarray}
-\frac{1}{2}\Biggl[ -\frac{3}{4} \stac{(0)}{K^2}
+ \stac{(0)}{\tilde K^\mu_\nu}  \stac{(0)}{\tilde K^\nu_\mu}    \Biggr] = \kappa^2 
\stac{(0)}{T}_{yy}
\end{eqnarray}
%
%=================================%
%===========<Equation>============%
%
\begin{eqnarray}
D_\nu \stac{(0)}{K^\nu_\mu} -D_\mu \stac{(0)}{K}=0
\end{eqnarray}
%
%=================================%
and
%===========<Equation>============%
%
\begin{eqnarray}
D^\alpha \tilde G_{y \alpha \alpha_1 \alpha_2 \alpha_3}=0. 
\end{eqnarray}
%
%=================================%
In the background the brane locations are $y=\stac{(0)}{\phi}_{\pm} =y_{(\pm)}$. 
The junction conditions are 
%===========<Equation>============%
%
\begin{eqnarray}
[\stac{(0)}{K}_{\mu\nu}-g_{\mu\nu} \stac{(0)}{K}]_{y=y_{(\pm)}} = \pm \frac{\kappa^2}{2}\gamma_{(\pm)} g_{\mu\nu}
\end{eqnarray}
%
%=================================%
and
%===========<Equation>============%
%
\begin{eqnarray}
\tilde G_{y \alpha_1 \alpha_2 \alpha_3 \alpha_4}(y_{(\pm)},x)= \pm \kappa^2 \gamma_{(\pm)} 
\epsilon_{\alpha_1 \alpha_2 \alpha_3 \alpha_4}(g_{(\pm)}). 
\end{eqnarray}
%
%=================================%

We find the solution of $\tilde G_5$ as  
%===========<Equation>============%
%
\begin{eqnarray}
\tilde G_{y \alpha_1 \alpha_2 \alpha_3 \alpha_4}(y,x)= 
\alpha  \epsilon_{\alpha_1 \alpha_2 \alpha_3 \alpha_4}(g),
\end{eqnarray}
%
%=================================%
where $\alpha$ will be determined by the junction condition. This is the solution in full order. 
Then we obtain 
%===========<Equation>============%
%
\begin{eqnarray}
\tilde G_{y \alpha_1 \alpha_2 \alpha_3 \alpha_4}(y,x)= 
\kappa^2 \gamma_{(+)}  \epsilon_{\alpha_1 \alpha_2 \alpha_3 \alpha_4}(g),\label{Gsol}
\end{eqnarray}
%
%=================================%
with 
%===========<Equation>============%
%
\begin{eqnarray}
\gamma_{(+)}=-\gamma_{(-)}.
\end{eqnarray}
%
%=================================%

Using 
%===========<Equation>============%
%
\begin{eqnarray}
\kappa^2 \stac{(0)}{T}_{yy} & = & \frac{1}{96}\tilde G_{y \alpha_1 \alpha_2 \alpha_3 \alpha_4} 
\tilde G_y^{~\alpha_1 \alpha_2 \alpha_3 \alpha_4}-\Lambda  \nonumber \\
 & = & -\frac{1}{4}\kappa^4 \gamma_{(+)}^2- \Lambda,
\label{eq:Tyy_zeroth}
\end{eqnarray}
%
%=================================%
and
%===========<Equation>============%
%
\begin{eqnarray}
\kappa^2 \Biggl({}^{(5)}T^\mu_\mu -\frac{4}{3} {}^{(5)}T^M_M \Biggr)^{(0)}
& = & -\frac{8}{3} \kappa^2 \stac{(0)}{T}_{yy},
\end{eqnarray}
%
%=================================%
the Hamiltonian constraint and the evolutional equation become
%===========<Equation>============%
%
\begin{eqnarray}
 \frac{3}{4}\stac{(0)}{K^2}=2\kappa^2 \stac{(0)}{T}_{yy}
\end{eqnarray}
%
%=================================%
and
%===========<Equation>============%
%
\begin{eqnarray}
\partial_y \stac{(0)}{K}=-\frac{1}{4}\stac{(0)}{K^2}+\frac{2}{3}\kappa^2 \stac{(0)}{T}_{yy}. 
\end{eqnarray}
%
%=================================%
Noting Eq. (\ref{eq:Tyy_zeroth}), we obtain the background solution as 
%===========<Equation>============%
%
\begin{eqnarray}
ds^2 =dy^2+a (y)^2 \gamma_{\mu\nu}dx^\mu dx^\nu
\end{eqnarray}
%
%=================================%
where 
%===========<Equation>============%
%
\begin{eqnarray}
a(y)=e^{-\frac{y}{\ell}}
\end{eqnarray}
%
%=================================%
and
%===========<Equation>============%
%
\begin{eqnarray}
\frac{1}{\ell} = \frac{1}{6}\kappa^2 \gamma_{(+)}.
\end{eqnarray}
%
%=================================%
Without loss of generality we can set $y_{(+)}=0$ and $y_{(-)}=y_0$. 
The extrinsic curvature is given by 
%===========<Equation>============%
%
\begin{eqnarray}
\stac{(0)}{K^\mu_\nu} =-\frac{1}{\ell} \delta^\mu_\nu.  
\end{eqnarray}
%
%=================================%

\subsection{First order}

In this subsection, we compute various quantities which are needed to derive the 
effective equation on the branes. For simplicity, we assume 
%===========<Equation>============%
%
\begin{eqnarray}
\frac{\phi_\pm - y_\pm }{\ell} \ll 1. 
\end{eqnarray}
%
%=================================%

First, we will obtain the solutions for form fields $\tilde F_3$ and $H_3$. We can solve
the equations including higher order because of the assumption,
$\tilde F_{\mu\nu\alpha}=0$ and $H_{\mu\nu\alpha}=0$. The equations which we will solve are 
%===========<Equation>============%
%
\begin{eqnarray}
\partial_y \tilde F_{y\mu\nu}-\frac{1}{2} \kappa^2 \gamma_{(+)} 
 H_{y\alpha\beta} \epsilon^{\alpha\beta}_{~~\mu\nu}=0
\end{eqnarray}
%
%=================================%
and
%===========<Equation>============%
%
\begin{eqnarray}
\partial_y H_{y \mu\nu}+ \frac{1}{2} \kappa^2 \gamma_{(+)} 
 \tilde F_{y\alpha\beta} \epsilon^{\alpha\beta}_{~~\mu\nu}=0.
\end{eqnarray}
%
%=================================%
The constraint equations are 
%===========<Equation>============%
%
\begin{eqnarray}
0=F_{\alpha_1 \alpha_2 \alpha_3} = \frac{1}{\kappa^2 \gamma_{(+)}} \epsilon_{\alpha_1 \alpha_2 \alpha_3 \mu}
D_\alpha H^{y \alpha \mu}
\end{eqnarray}
%
%=================================%
and
%===========<Equation>============%
%
\begin{eqnarray}
0=H_{\alpha_1 \alpha_2 \alpha_3} = -\frac{1}{\kappa^2 \gamma_{(+)}} \epsilon_{\alpha_1 \alpha_2 \alpha_3 \mu}
D_\alpha F^{y \alpha \mu}. 
\end{eqnarray}
%
%=================================%
%From them we see that $\phi$ must be constant. See the Appendix A for the detail. 

The general solutions are 
%===========<Equation>============%
%
\begin{eqnarray}
H_{y \mu\nu} = a^{-6} \alpha_{\mu\nu} + a^6 \beta_{\mu\nu}
\end{eqnarray}
%
%=================================%
and
%===========<Equation>============%
%
\begin{eqnarray}
\tilde F_{y\mu\nu} = 
\frac{1}{2} \epsilon_{\mu\nu}^{~~~\alpha\beta} (a^{-6}\alpha_{\alpha \beta}
-a^6 \beta_{\alpha \beta}).
\end{eqnarray}
%
%=================================%
The junction condition implies the relation 
%===========<Equation>============%
%
\begin{eqnarray}
e^{\frac{6}{\ell} y_{\pm}} \alpha_{\mu\nu} + e^{-\frac{6}{\ell}  y_\pm} \beta_{\mu\nu}
=\pm \kappa^2 \beta_{(\pm)}{\cal F}^{(\pm)}_{\mu\nu}
\end{eqnarray}
%
%=================================%
and
%===========<Equation>============%
%
\begin{eqnarray}
e^{\frac{6}{\ell} y_{\pm}} \alpha_{\mu\nu} - e^{-\frac{6}{\ell}  y_\pm} \beta_{\mu\nu}
=\pm \kappa^2 \gamma_{(\pm)}{\cal F}^{(\pm)}_{\mu\nu}. 
\end{eqnarray}
%
%=================================%
That is, 
%===========<Equation>============%
%
\begin{eqnarray}
\alpha_{\mu\nu}(x)= \pm \kappa^2 e^{-\frac{6}{\ell} y_{\pm}} (\beta_{(\pm)}+\gamma_{(\pm)} ) 
{\cal F}^{(\pm)}_{\mu \nu} \label{alphamunu}
\end{eqnarray}
%
%=================================%
and
%===========<Equation>============%
%
\begin{eqnarray}
\beta_{\mu\nu}(x)= \pm \kappa^2 e^{\frac{6}{\ell} y_{\pm}} (\beta_{(\pm)}-\gamma_{(\pm)} ) 
{\cal F}^{(\pm)}_{\mu \nu} \label{betamunu}
\end{eqnarray}
%
%=================================%
Since the solution for $\alpha_{\mu\nu} (x) $ and $\beta_{\mu\nu} (x)$ have two different forms
as above, we have the following relation
%===========<Equation>============%
%
\begin{eqnarray}
{\cal F}^{(-)}_{\mu\nu} & = &  -e^{- \frac{6}{\ell} (y_+ -y_-)} 
\frac{\beta_{(+)} + \gamma_{(+)}}{\beta_{(-)} + \gamma_{(-)}} 
{\cal F}^{(+)}_{\mu\nu} \nonumber \\
& = & -e^{\frac{6}{\ell} (y_+ -y_-)} \frac{\beta_{(+)} -\gamma_{(+)}}{\beta_{(-)} - \gamma_{(-)}} 
{\cal F}^{(+)}_{\mu\nu}. \label{ffrelation}
\end{eqnarray}
%
%=================================%
Moreover, we see 
%===========<Equation>============%
%
\begin{eqnarray}
e^{\frac{12}{\ell}  (y_+-y_-)}= \frac{\beta_{(+)} + \gamma_{(+)}}{\beta_{(+)} - \gamma_{(+)}}
\frac{\beta_{(-)}-\gamma_{(-)}}{\beta_{(-)} + \gamma_{(-)}}.
\label{eq:relation_beta-gamma}
\end{eqnarray}
%
%=================================%
Finally we obtain 
%===========<Equation>============%
%
\begin{eqnarray}
H_{y \mu\nu} = \kappa^2 \Biggl[ \beta_{(+)}  {\rm cosh} \Bigl( \frac{6}{\ell} y \Bigr)
+ \gamma_{(+)} 
{\rm sinh} \Bigl( \frac{6}{\ell} y \Bigr) \Biggr] {\cal F}^{(+)}_{\mu\nu} \label{Hsol}
\end{eqnarray}
%
%=================================%
and
%===========<Equation>============%
%
\begin{eqnarray}
\tilde F_{y \mu\nu} =   \frac{\kappa^2}{2} \Biggl[ \beta_{(+)}  
{\rm sinh} \Bigl( \frac{6}{\ell} y \Bigr)
+ \gamma_{(+)} 
{\rm cosh}\Bigl(\frac{6}{\ell} y  \Bigr) \Biggl] \epsilon_{\mu\nu}^{~~\alpha\beta}{\cal F}^{(+)}_{\alpha \beta}.
\label{Fsol}
\end{eqnarray}
%
%=================================%

Since the form fields can contribute to the effective equation in the quadratic term, 
the lowest order solution will be half order and then 
%===========<Equation>============%
%
\begin{eqnarray}
\stac{(1/2)}{H}_{y \mu\nu}= \kappa^2 \gamma_{(+)}a^{-6}\stac{(+)}{{\cal F}_{\mu\nu}}
\end{eqnarray}
%
%=================================%
and
%===========<Equation>============%
%
\begin{eqnarray}
\stac{(1/2)}{\tilde F}_{y \mu\nu}= \frac{1}{2}\kappa^2 \gamma_{(+)}a^{-6}
\epsilon_{\mu\nu}^{~~\alpha\beta}\stac{(+)}{{\cal F}_{\alpha\beta}}. 
\end{eqnarray}
%
%=================================%
From Eq. (\ref{ffrelation}) 
%===========<Equation>============%
%
\begin{eqnarray}
\stac{(-)}{{\cal F}_{\mu\nu}}=a_0^{-6}\stac{(+)}{{\cal F}_{\mu\nu}}
\end{eqnarray}
%
%=================================%
and then 
%===========<Equation>============%
%
\begin{eqnarray}
T_{\mu\nu}^{(-)}= a^{-14}_0 T_{\mu\nu}^{(+)}
\end{eqnarray}
%
%=================================%
hold. Here $a_0=e^{-\frac{y_0}{\ell}}$. 

Now we are ready to solve the extrinsic curvature. We first solve the traceless part which 
follows the evolutional equation
%===========<Equation>============%
%
\begin{eqnarray}
\partial_y \stac{(1)}{\tilde K^\mu_\nu} ={}^{(4)}\stac{(1)}{\tilde R^\mu_\nu} -\stac{(0)}{K} 
\stac{(1)}{\tilde K^\mu_\nu} -\kappa^2 
\Bigl({}^{(5)}T^\mu_\nu -\frac{1}{4}\delta^\mu_\nu {}^{(5)}T^\alpha_\alpha  \Bigr)^{(1)}.
\end{eqnarray}
%
%=================================%
Then we obtain the first order solution of $\stac{(1)}{{\tilde K}^\mu_\nu} $
%===========<Equation>============%
%
\begin{eqnarray}
\stac{(1)}{{\tilde K}^\mu_\nu}(y,x) & = &  -\frac{\ell}{2a^2}{}^{(4)} \tilde R^\mu_\nu (h)
-\frac{1}{2}\kappa^2 \gamma_{(+)}a^{-16}T^{\mu (+)}_\nu (h) \nonumber \\
& & +\frac{\stac{(1)}{\chi^\mu_\nu }(x)}{a^4},
\end{eqnarray}
%
%=================================%
where $\stac{(1)}{\chi^\mu_\nu }(x) $ is the constant of integration.  

In this order the junction condition for $\stac{(1)}{{\tilde K}^\mu_\nu}(y,x) $ on the branes is 
given by 
%===========<Equation>============%
%
\begin{eqnarray}
\stac{(1)}{{\tilde K}^\mu_\nu}(y_\pm,x) & = & \mp \frac{\kappa^2}{2}\gamma_{(\pm)}T^{\mu (\pm)}_\nu
+\Bigl[ D^\mu D_\nu \phi_\pm \nonumber \\
& & +\frac{1}{\ell} D^\mu \phi_\pm D_\nu \phi_\pm \Bigr]_{\rm traceless}. 
\end{eqnarray}
%
%=================================%
On $D_+$ brane, therefore, we obtain 
%===========<Equation>============%
%
\begin{eqnarray}
& & \Bigl( {\cal D}^\mu {\cal D}_\nu \phi_+ +\frac{1}{\ell}{\cal D}^\mu \phi_+ {\cal D}_\nu \phi_+ 
\Bigr)_{\rm traceless} \nonumber \\
& & ~~ = 
-\frac{\ell}{2} {}^{(4)}\tilde R^\mu_\nu (h) + \stac{(1)}{\chi^\mu_\nu}(x) \label{1sttlp}
\end{eqnarray}
%
%=================================%
and, on $D_-$ brane, 
%===========<Equation>============%
%
\begin{eqnarray}
& & a_0^{-2}\Bigl( {\cal D}^\mu {\cal D}_\nu \phi_- +\frac{1}{\ell}{\cal D}^\mu \phi_- {\cal D}_\nu \phi_-
\Bigr)_{\rm traceless}\nonumber \\
& & ~~=-a_0^{-2}\frac{\ell}{2} {}^{(4)}\tilde R^\mu_\nu (h)+ a_0^{-4}\stac{(1)}{\chi^\mu_\nu}(x),
\label{1sttlm}
\end{eqnarray}
%
%=================================%
where ${\cal D}_\mu$ is the covariant derivative with respect to $h_{\mu\nu}$. 
Eliminating $\stac{(1)}{\chi^\mu_\nu}(x)$ by using Eqs. (\ref{1sttlp}) and (\ref{1sttlm}), 
we see 
%===========<Equation>============%
%
\begin{eqnarray}
& & \Bigl( {\cal D}^\mu {\cal D}_\nu \phi_+ +\frac{1}{\ell}{\cal D}^\mu \phi_+ {\cal D}_\nu \phi_+ 
\Bigr)_{\rm traceless} \nonumber \\
& & ~~-a^{2}_0 
\Bigl( {\cal D}^\mu {\cal D}_\nu \phi_- +\frac{1}{\ell}{\cal D}^\mu \phi_- {\cal D}_\nu \phi_-
\Bigr)_{\rm traceless} \nonumber \\
& & ~~~~= -\frac{\ell}{2}(1-a^2_0) {}^{(4)}\tilde R^\mu_\nu (h). \label{1sttl}
\end{eqnarray}
%
%=================================%

Let us compute the trace part of the extrinsic curvature. From the Hamiltonian constraint
%===========<Equation>============%
%
\begin{eqnarray}
-\frac{1}{2a^2}{}^{(4)}R(h)+\frac{3}{4}\stac{(0)}{K} \stac{(1)}{K}= \kappa^2 [{}^{(5)}T_{yy}]^{(1)},
\end{eqnarray}
%
%=================================%
noting that $T_{yy}$ vanishes at the first order, we obtain
%===========<Equation>============%
%
\begin{eqnarray}
\stac{(1)}{K} (y,x) = -\frac{\ell}{6a^2}{}^{(4)}R(h).
\end{eqnarray}
%
%=================================%
Then the junction condition for $\stac{(1)}{K}$ on the branes, 
%===========<Equation>============%
%
\begin{eqnarray}
\stac{(1)}{K}(y_\pm, x) & = &  \mp \frac{2}{3} \kappa^2 (\beta_{(\pm)}-\gamma_{(\pm)})
+\stac{(\pm)}{D^2} \phi_\pm \nonumber \\
& & -\frac{1}{\ell} (\stac{(\pm)}{D} \phi_\pm )^2, 
\end{eqnarray}
%
%=================================%
implies 
%===========<Equation>============%
%
\begin{eqnarray}
& & -\frac{2}{3}\kappa^2 ( \beta_{(+)}-\gamma_{(+)}) +{\cal D}^2 \phi_+ 
-\frac{1}{\ell}({\cal D}\phi_+ )^2 \nonumber \\
& & ~~~~~=-\frac{\ell}{6}{}^{(4)}R(h) \label{1sttrp}
\end{eqnarray}
%
%=================================%
and
%===========<Equation>============%
%
\begin{eqnarray}
& & \frac{2}{3}\kappa^2 ( \beta_{(-)}-\gamma_{(-)}) + a_0^{-2} \Bigl( {\cal D}^2 \phi_- 
-\frac{1}{\ell}({\cal D}\phi_- )^2\Bigr) \nonumber \\
& & ~~~~~=-\frac{\ell}{6a_0^2}{}^{(4)}R(h). \label{1sttrm}
\end{eqnarray}
%
%=================================%
From the above two equations, we can construct the following equation
%===========<Equation>============%
%
\begin{eqnarray}
& & -\frac{2}{3}\kappa^2 (\beta_{(+)}-\gamma_{(+)})-\frac{2}{3}\kappa^2 a_0^4 (\beta_{(-)}-\gamma_{(-)})
\nonumber \\
& & ~~+{\cal D}^2 \phi_+ -\frac{1}{\ell} ({\cal D}\phi_+ )^2
-a_0^2 \Bigl( {\cal D}^2 \phi_- -\frac{1}{\ell} ({\cal D}\phi_- )^2 \Bigr) \nonumber \\
& & ~~~~
= -\frac{\ell}{6}(1-a^2_0){}^{(4)}R(h). \label{1sttr}
\end{eqnarray}
%
%=================================%
From Eqs. (\ref{1sttl}) and (\ref{1sttr}) we obtain the first order effective Einstein equation 
with respect to the metric $h_{\mu\nu}$ as 
\begin{widetext}
%===========<Equation>============%
%
\begin{eqnarray}
(1-a_0^2) {}^{(4)}G_{\mu\nu}(h)
& = & -\frac{\kappa^2}{\ell} \Bigl[\beta_{(+)}-\gamma_{(+)}+a_0^4 (\beta_{(-)}-\gamma_{(-)})  
\Bigr]h_{\mu\nu} \nonumber \\
& & -\frac{2}{\ell} \Bigl[
{\cal D}_\mu {\cal D}_\nu \phi_+ -h_{\mu\nu} {\cal D}^2 \phi_+ 
+\frac{1}{\ell} \Bigl({\cal D}_\mu \phi_+ {\cal D}_\nu \phi_+ + \frac{1}{2}h_{\mu\nu} 
({\cal D} \phi_+)^2 \Bigr) \nonumber \\
& & -a_0^2 \Bigl( 
{\cal D}_\mu {\cal D}_\nu \phi_- -h_{\mu\nu} {\cal D}^2 \phi_- 
+\frac{1}{\ell} \Bigl({\cal D}_\mu \phi_- {\cal D}_\nu \phi_- +\frac{1}{2}h_{\mu\nu} 
({\cal D} \phi_-)^2 \Bigr) \Bigr) \Bigr].
\end{eqnarray}
%
%=================================%
This is not the end of this section because $h_{\mu\nu}$ is not induced metric 
$g_{(+) \mu\nu}= a^2(\phi_+) h_{\mu\nu}+\partial_\mu \phi_+ \partial_\nu \phi_+$
on $D_+$ brane. Then we must rewrite the above effective equation with respect to 
the induced metric $g_{(+) \mu\nu}$ 
%===========<Equation>============%
%
\begin{eqnarray}
(1-a_0^2) {}^{(4)}G_{\mu\nu}(g_{(+)})
& = &   -\frac{\kappa^2}{\ell} \Bigl[\beta_{(+)}-\gamma_{(+)}+a_0^4 (\beta_{(-)}-\gamma_{(-)})  
\Bigr]g_{(+)\mu\nu} \nonumber \\
& &
+ \frac{2}{\ell}a_0^2 \Bigl[\stac{(+)}{D_\mu} \stac{(+)}{D_\nu}d 
-g_{(+)\mu\nu} \stac{(+)}{D^2}d
+\frac{1}{\ell} \Bigl \lbrace \stac{(+)}{D_\mu} d \stac{(+)}{D_\nu}d 
+\frac{1}{2} g_{(+)\mu\nu} (\stac{(+)}{D} d)^2 \Bigr\rbrace
\Bigr].
\end{eqnarray}
%
%=================================%
\end{widetext}
where $d(x)= \phi_- (x) - \phi_+ (x)$ is the proper distance between the two branes and
$\stac{(+)}{D_\mu}$ is a covariant derivative on $D_+$ brane. 
The equation for the radion can be derived from Eqs. (\ref{1sttrp}) and (\ref{1sttrm}) as 
%===========<Equation>============%
%
\begin{eqnarray}
& & a^2_0 \Bigl(- \stac{(+)}{D^2} d +\frac{1}{\ell} (\stac{(+)}{D}d )^2  \Bigr)
-\frac{2}{3}\kappa^2 (\beta_{(+)}-\gamma_{(+)}) \nonumber \\
& & ~~-\frac{2}{3}\kappa^2 a_0^2 
(\beta_{(-)}-\gamma_{(-)}) =0.
\end{eqnarray}
%
%=================================%

As expected, the Maxwell field cannot be a source for the gravity on the branes at this order. 
This result is consistent with one obtained in the previous studies \cite{SKOT,OSKH} because 
the background spacetime satisfies BPS condition. To see the non-zero gravitational coupling, the 
next order corrections will be important.

\subsection{Second order}

Let us consider the second order perturbations. The order of the form fields which can 
contribute to the second order corrections to the effective theory 
is $(\beta_{(\pm)}-\gamma_{(\pm)}){\cal F}_{\mu\nu}$. 
Indeed, the solution of the form fields at this order are obtain as 
%===========<Equation>============%
%
\begin{eqnarray}
\stac{(3/2)}{H}_{y\mu\nu} = \frac{1}{2}\kappa^2 (\beta_{(+)}-\gamma_{(+)})
(a^{-6}+a^6 ) \stac{(+)}{{\cal F}_{\mu\nu}}
\end{eqnarray}
%
%=================================%
and
%===========<Equation>============%
%
\begin{eqnarray}
\stac{(3/2)}{\tilde F_{y\mu\nu}}= \frac{1}{4}\kappa^2 (\beta_{(+)}-\gamma_{(+)})
(a^{-6}-a^6 ) \epsilon_{\mu\nu}^{~~\alpha\beta} \stac{(+)}{{\cal F}_{\alpha\beta}}. 
\end{eqnarray}
%
%=================================%

The evolutional equation for the traceless part of the extrinsic curvature is 
%===========<Equation>============%
%
\begin{eqnarray}
\partial_y \stac{(2)}{{\tilde K}^\mu_\nu} 
& = &  [\tilde R^\mu_\nu]^{(2)}-\stac{(1)}{K}\stac{(1)}{{\tilde K}^\mu_\nu} 
-\stac{(0)}{K}\stac{(2)}{{\tilde K}^\mu_\nu} \nonumber \\
& & -\kappa^2 \Bigl( {}^{(5)}T^\mu_\nu -\frac{1}{4} \delta^\mu_\nu {}^{(5)}T^\alpha_\alpha \Bigr). 
\end{eqnarray}
%
%=================================%
To compute the right-hand side we need the first order deviation from the seed metric in the bulk 
%===========<Equation>============%
%
\begin{eqnarray}
g_{\mu\nu}= a ^2(h_{\mu\nu}+ \stac{(1)}{g}_{\mu\nu}). 
\end{eqnarray}
%
%=================================%
Since we know the solution to the extrinsic curvature at the first order, we have the 
equation for $\stac{(1)}{g}_{\mu\nu}$
%===========<Equation>============%
%
\begin{eqnarray}
\stac{(1)}{g}_{\mu\nu} & = & \frac{\ell^2}{2}(1-a^{-2})
\Bigl( {}^{(4)}R_{\mu\nu}-\frac{1}{6}h_{\mu\nu}{}^{(4)}R \Bigr) \nonumber \\
& & +\frac{\ell^2}{4}(a^{-4}-1){}^{(4)}\tilde R_{\mu\nu}
+\frac{3}{8}(1-a^{-16})T^{(+)}_{\mu\nu} + \cdots \nonumber  \\
& = & \frac{\kappa^2}{6}\ell (1-a^{-2})\frac{1+\delta a_0^4 }{1-a_0^2}(\beta_{(+)}-\gamma_{(+)}) h_{\mu\nu} 
\nonumber \\
& & +\frac{3}{8}(1-a^{-16})T^{(+)}_{\mu\nu} +\cdots,  
\end{eqnarray}
%
%=================================%
where 
%===========<Equation>============%
%
\begin{eqnarray}
\delta = \frac{\beta_{(-)}-\gamma_{(-)}}{\beta_{(+)}-\gamma_{(+)}}. 
\end{eqnarray}
%
%=================================%
In the above we used the effective Einstein equation at the first order. Then 
we can compute the second order part of the Ricci tensor ${}^{(4)}R$ 
%===========<Equation>============%
%
\begin{eqnarray}
[{}^{(4)}R^\mu_\nu]^{(2)} & = &  \frac{1}{6}\Bigl( \frac{1+\delta a_0^4}{1-a_0^2} \Bigr)^2 (a^{-4}-a^{-2})
\kappa^4 (\gamma_{(+)}-\beta_{(+)})^2 \delta^\mu_\nu
\nonumber \\
& & +\frac{3}{8} \frac{1+\delta a_0^2}{1-a_0^2}(a^{-18}-a^{-2})\frac{\kappa^2}{\ell}(\beta_{(+)}-\gamma_{(+)})T^{\mu (+)}_\nu \nonumber \\
& & -\frac{3}{16}(a^{-18}-a^{-2})({\cal D}_\alpha {\cal D}_\nu T^{\mu (+)}_\nu 
\nonumber \\
& & +{\cal D}_\alpha {\cal D}^\mu T^{(+)}_{\nu\alpha}-{\cal D}^2T^{\mu (+)}_\nu ). 
\end{eqnarray}
%
%=================================%
Using this and the solutions of $\stac{(3/2)}{H}_{y\mu\nu}$, $\stac{(3/2)}{\tilde F_{y\mu\nu}}$, 
we see 
\begin{widetext}
%===========<Equation>============%
%
\begin{eqnarray}
-\kappa^2 [{}^{(5)}T^\mu_\nu ]^{(2)}_{\rm traceless} & = &  
-\frac{1}{2a^4} \Bigl[ 
-\stac{(1)}{g^{\mu\beta}} \stac{(1/2)}{H_{\beta y \alpha}} \stac{(1/2)}{H_{\nu y}^{~~\alpha}}
-\stac{(1)}{g^{\alpha \rho}} \stac{(1/2)}{H_{\beta y \alpha}} \stac{(1/2)}{H_{\nu y \rho}} h^{\mu \beta}
+\stac{(1/2)}{H^\mu_{~y\alpha}} \stac{(3/2)}{H_{\nu y}^{~~\alpha}}
+\stac{(3/2)}{H^\mu_{~y\alpha}} \stac{(1/2)}{H_{\nu y}^{~~\alpha}}
+ (H \to \tilde F) \Bigr]_{\rm traceless} \nonumber \\
& = & \kappa^4 \gamma_{(+)}(\gamma_{(+)}-\beta_{(+)}) \Biggl[ 2\frac{1+\delta a_0^4}{1-a_0^2}
a^{-18}-\Bigl(2\frac{1+\delta a_0^4}{1-a_0^2} -1 \Bigr)a^{-16} \Biggr] T^{\mu (+)}_\nu. 
\end{eqnarray}
%
%=================================%
Then the evolutional equation in the second order becomes
%===========<Equation>============%
%
\begin{eqnarray}
\partial_y \stac{(2)}{{\tilde K}^\mu_\nu} & = & \frac{4}{\ell} \stac{(2)}{{\tilde K}^\mu_\nu}
+\kappa^4 \gamma_{(+)}(\gamma_{(+)}-\beta_{(+)})
\Biggl[ \frac{109}{48}\frac{1+\delta a_0^4}{1-a_0^2} a^{-18}
-\Bigl( 2\frac{1+\delta a_0^4}{1-a_0^2}-1  \Bigr)a^{-16}
+\frac{1}{16}\frac{1+\delta a_0^4}{1-a_0^2}a^{-2}
\Biggr]T^{\mu (+)}_\nu (h) \nonumber \\
& & -\frac{3}{16}(a^{-18}-a^{-2})({\cal D}_\alpha {\cal D}_\nu T^{\mu (+)}_\nu 
+{\cal D}_\alpha {\cal D}^\mu T^{(+)}_{\nu\alpha}
-{\cal D}^2T^{\mu (+)}_\nu ). 
\end{eqnarray}
%
%=================================%
Now the solution is given by 
%===========<Equation>============%
%
\begin{eqnarray}
\stac{(2)}{{\tilde K}^\mu_\nu} & = &  
-\frac{3 \ell}{16} \Bigl(\frac{1}{14}a^{-18}+\frac{1}{2}a^{-2}\Bigr)
\Bigl({\cal D}_\alpha {\cal D}_\nu T^{\mu (+)}_\alpha +{\cal D}_\alpha {\cal D}^\mu 
T^{(+)}_{\nu\alpha}-{\cal D}^2 T^{\mu (+)}_\nu \Bigr) \nonumber \\
& & +\kappa^4 \ell \gamma_{(+)} (\gamma_{(+)}-\beta_{(+)}) 
\Biggl[ 
\frac{109}{48 \cdot 14} \frac{1+\delta a_0^4}{1-a_0^2} a^{-18}
-\frac{1}{12}\Bigl( 2\frac{1+\delta a_0^4}{1-a_0^2}-1  \Bigr)a^{-16}
-\frac{1}{32}\frac{1+\delta a_0^4}{1-a_0^2}a^{-2}
\Biggr]T^{\mu (+)}_\nu 
+\frac{\stac{(2)}{\chi^\mu_\nu}(x)}{a^4}. 
\end{eqnarray}
%
%=================================%
Up to the second order the traceless part of the extrinsic curvature becomes 
%===========<Equation>============%
%
\begin{eqnarray}
\tilde K^\mu_\nu & = &  \stac{(1)}{{\tilde K}^\mu_\nu} +\stac{(2)}{{\tilde K}^\mu_\nu}
= -\frac{\ell}{2a^2}{}^{(4)} \tilde R^\mu_\nu (h) -\frac{1}{2}\kappa^2 \gamma_{(+)}
a^{-16}T^{\mu (+)}_\nu (h) \nonumber \\
& & -\frac{3 \ell}{16} \Bigl(\frac{1}{14}a^{-18}+\frac{1}{2}a^{-2}\Bigr)
\Bigl({\cal D}_\alpha {\cal D}_\nu T^{\mu (+)}_\alpha +{\cal D}_\alpha {\cal D}^\mu 
T^{(+)}_{\nu\alpha}-{\cal D}^2 T^{\mu (+)}_\nu \Bigr) \nonumber \\
& & +\kappa^4 \ell \gamma_{(+)} (\gamma_{(+)}-\beta_{(+)}) 
\Biggl[ 
\frac{109}{48 \cdot 14} \frac{1+\delta a_0^4}{1-a_0^2} a^{-18}
-\frac{1}{12}\Bigl( 2\frac{1+\delta a_0^4}{1-a_0^2}-1  \Bigr)a^{-16}
-\frac{1}{32}\frac{1+\delta a_0^4}{1-a_0^2}a^{-2}
\Biggr]T^{\mu (+)}_\nu 
+\frac{\chi^\mu_\nu(x)}{a^4}
\end{eqnarray}
%
%=================================%
where $\chi^\mu_\nu (x) =  \stac{(1)}{\chi^\mu_\nu}(x)+ \stac{(2)}{\chi^\mu_\nu}(x)  $. 
Then the junction conditions give us 
%===========<Equation>============%
%
\begin{eqnarray}
\Bigl( {\cal D}^\mu {\cal D}_\nu \phi_+ 
+\frac{1}{\ell}{\cal D}^\mu \phi_+ {\cal D}_\nu \phi_+   \Bigr)_{\rm traceless}
& = & -\frac{\ell}{2}{}^{(4)}R(h) +\kappa^2 \Bigl(\frac{109}{112}-1-\frac{3}{16} \Bigr)
(\gamma_{(+)}-\beta_{(+)}) \frac{1+\delta a_0^4}{1-a_0^2}
T^{\mu (+)}_\nu (h) \nonumber \\
& &  -\frac{3}{28}\ell \Bigl({\cal D}_\alpha {\cal D}_\nu T^{\mu (+)}_\alpha +{\cal D}_\alpha {\cal D}^\mu 
T^{(+)}_{\nu\alpha}-{\cal D}^2 T^{\mu (+)}_\nu \Bigr) +\chi^\mu_\nu (x) \label{2ndtlp}
\end{eqnarray}
%
%=================================%
and
%===========<Equation>============%
%
\begin{eqnarray}
a_0^{-2} \Bigl( {\cal D}^\mu {\cal D}_\nu \phi_-
+\frac{1}{\ell}{\cal D}^\mu \phi_- {\cal D}_\nu \phi_-   \Bigr)_{\rm traceless}
& = & -\frac{\ell}{2a_0^2}{}^{(4)}R(h) 
-\frac{1}{2}\kappa^2 (\gamma_{(+)}+\beta_{(-)}) a^{-16}_0 T^{\mu (+)}_\nu (h) \nonumber \\
& & +\kappa^2 (\gamma_{(+)}-\beta_{(+)})  \Biggl[
\frac{109}{112} \frac{1+\delta a_0^4}{1-a_0^2} a_0^{-18}
-\frac{1}{2} \Bigl(2 \frac{1+\delta a_0^4}{1-a_0^2} -1-\delta  \Bigr) a_0^{-16} \nonumber \\
& & -\frac{3}{16}\frac{1+\delta a_0^4}{1-a_0^2}a_0^{-2} 
\Biggr]T^{\mu (+)}_\nu 
\nonumber \\
& & -\frac{3}{16} \ell \Bigl(\frac{1}{14}a^{-18}_0+\frac{1}{2}a^{-2}_0 \Bigr)
\Bigl({\cal D}_\alpha {\cal D}_\nu T^{\mu (+)}_\alpha +{\cal D}_\alpha {\cal D}^\mu T^{(+)}_{\nu\alpha}-{\cal D}^2 T^{\mu (+)}_\nu \Bigr)
\nonumber \\
& & +\chi^\mu_\nu a_0^{-4}. \label{2ndtlm}
\end{eqnarray}
%
%=================================%
Using Eqs. (\ref{2ndtlp}) and (\ref{2ndtlm}), we can eliminate $\chi^\mu_\nu$ as 
%===========<Equation>============%
%
\begin{eqnarray}
& & \Bigl( {\cal D}^\mu {\cal D}_\nu \phi_+ +\frac{1}{\ell}{\cal D}^\mu \phi_+ {\cal D}_\nu \phi_+ 
\Bigr)_{\rm traceless}
-a^2_0 \Bigl( {\cal D}^\mu {\cal D}_\nu \phi_- +\frac{1}{\ell}{\cal D}^\mu \phi_-{\cal D}_\nu \phi_-
\Bigr)_{\rm traceless} \nonumber \\
& & ~~~~
=-\frac{\ell}{2}(1-a_0^2) {}^{(4)} \tilde R^\mu_\nu (h) 
+\kappa^2 (\gamma_{(+)}+\beta_{(-)})
\Biggl[ \frac{109}{112}\frac{1+\delta a_0^4}{1-a_0^2}(1-a_0^{-14})
-\frac{3}{16}(1+\delta a_0^4) -\frac{1}{2}(1+\delta )a_0^{-12}
\Biggr]T^{\mu (+)}_\nu (h) \nonumber \\
& & ~~~~~~-\frac{3}{16}
\Bigl( \frac{1}{14}(1-a^{-14}_0)+\frac{1}{2}(1-a_0^2)  \Bigr)
\Bigl({\cal D}_\alpha {\cal D}_\nu T^{\mu (+)}_\alpha +{\cal D}_\alpha {\cal D}^\mu T^{(+)}_{\nu\alpha}-{\cal D}^2 T^{\mu (+)}_\nu \Bigr). 
\label{2ndtl}
\end{eqnarray}
%
%=================================%
\end{widetext}
This is the traceless part of the effective gravitational equation on $D_+$ brane. 

From the second order Hamiltonian equation 
%===========<Equation>============%
%
\begin{eqnarray}
-\frac{1}{2}({}^{(4)}R)^{(2)} +\frac{3}{4}\stac{(0)}{K} \stac{(2)}{K}+\frac{3}{8} \stac{(1)}{K^2}
= \kappa^2 (T_{yy})^{(2)}, 
\end{eqnarray}
%
%=================================%
where we neglected the contribution from the first-order traceless extrinsic curvatures
because of Eq. (\ref{eq:hierarchy}), we obtain 
%===========<Equation>============%
%
\begin{eqnarray}
\stac{(2)}{K} & = & \frac{\ell}{18}\kappa^4 (\gamma_{(+)}-\beta_{(+)})^2 
\Bigl(\frac{1+\delta a_0^4}{1-a_0^2} \Bigr)^2 (-a^{-4}+2a^{-2})
\nonumber \\
& & -\frac{1}{2}\kappa^2 (\beta_{(+)}-\gamma_{(+)}) \stac{(+)}{{\cal F}_{\mu\nu}}
     \stac{(+)}{{\cal F}^{\mu\nu}}(h)a^{-4}
\end{eqnarray}
%
%=================================%
Together with the result at the first order 
\begin{widetext}
%===========<Equation>============%
%
\begin{eqnarray}
K = \stac{(1)}{K} +  \stac{(2)}{K}
= -\frac{4}{\ell}-\frac{\ell}{6a^2} {}^{(4)}R(h) 
+\frac{\ell}{18}\kappa^4 (\gamma_{(+)}-\beta_{(+)})^2 
\Bigl(\frac{1+\delta a_0^4}{1-a_0^2} \Bigr)^2 (-a^{-4}+2a^{-2})
-\frac{1}{2}\kappa^2 (\beta_{(+)}-\gamma_{(+)}) \stac{(+)}{{\cal F}_{\mu\nu}}
 \stac{(+)}{{\cal F}^{\mu\nu}}(h)a^{-4}. 
\end{eqnarray}
%
%=================================%
Then the junction conditions imply 
%===========<Equation>============%
%
\begin{eqnarray}
-\frac{2}{3}\kappa^2 (\beta_{(+)}-\gamma_{(+)})
+{\cal D}^2 \phi_+ - \frac{1}{\ell}({\cal D}\phi_+)^2 
=-\frac{\ell}{6}{}^{(4)}R(h)+\frac{\ell}{18}\kappa^4 (\gamma_{(+)}-\beta_{(+)})^2 
\Bigl(\frac{1+\delta a_0^4}{1-a_0^2} \Bigr)^2
-\frac{1}{2}\kappa^2 (\beta_{(+)}-\gamma_{(+)}) \stac{(+)}{{\cal F}_{\mu\nu}}
 \stac{(+)}{{\cal F}^{\mu\nu}}(h)
\end{eqnarray}
%
%=================================%
and
%===========<Equation>============%
%
\begin{eqnarray}
\frac{2}{3}\kappa^2 (\beta_{(-)}-\gamma_{(-)})
+a^{-2}_0 \Bigl({\cal D}^2 \phi_- -\frac{1}{\ell}({\cal D}\phi_-)^2    \Bigr) 
 & = & -\frac{\ell}{6a_0^2}{}^{(4)}R(h)
+\frac{\ell}{18}\kappa^4 (\gamma_{(+)}-\beta_{(+)})^2 \Bigl(\frac{1+\delta a_0^4}{1-a_0^2} \Bigr)^2
(2a_0^{-2}-a_0^{-4})
\nonumber \\
& & -\frac{1}{2}\kappa^2 (\beta_{(+)}-\gamma_{(+)}) \stac{(+)}{{\cal F}_{\mu\nu}}
 \stac{(+)}{{\cal F}^{\mu\nu}}(h)a^{-4}_0. 
\end{eqnarray}
%
%=================================%
From these we see 
%===========<Equation>============%
%
\begin{eqnarray}
& & -\frac{2}{3}\kappa^2  (\beta_{(+)}-\gamma_{(+)})
-a_0^4 \frac{2}{3}\kappa^2 (\beta_{(-)}-\gamma_{(-)})
+{\cal D}^2 \phi_+ -\frac{1}{\ell}({\cal D}\phi_+)^2 
-a^{2}_0 \Bigl({\cal D}^2 \phi_- -\frac{1}{\ell}({\cal D}\phi_-)^2    \Bigr) \nonumber \\
 & & ~~~~=  -\frac{\ell}{6}(1-a_0^2){}^{(4)}R(h) +\frac{\ell}{9}
\kappa^2 (\gamma_{(+)}-\beta_{(+)})^2 \frac{(1+\delta a_0^4)^2}{1-a_0^2}
\label{2ndtr}
\end{eqnarray}
%
%=================================%
holds. Using Eqs. (\ref{2ndtl}) and (\ref{2ndtr}) we finally obtain the 
effective Einstein equation with respect to the metric $h_{\mu\nu}$
%===========<Equation>============%
%
\begin{eqnarray}
(1-a^2_0) G_{\mu\nu}(h) 
 & = &  -\frac{2}{\ell} \Bigl[ 
{\cal D}_\mu {\cal D}_\nu \phi_+ -h_{\mu\nu}{\cal D}^2 \phi_+ 
-a_0^2 ( {\cal D}_\mu {\cal D}_\nu \phi_- -h_{\mu\nu}{\cal D}^2 \phi_- ) \nonumber \\
& & +\frac{1}{\ell} \Bigl\lbrace
{\cal D}_\mu \phi_+ {\cal D}_\nu \phi_+ +\frac{1}{2} h_{\mu\nu}({\cal D} \phi_+ )^2 
-a_0^2 ({\cal D}_\mu \phi_- {\cal D}_\nu \phi_- +\frac{1}{2} h_{\mu\nu}({\cal D} \phi_- )^2   )
\Bigr\rbrace \Bigl] \nonumber \\
& & -\kappa^2 \ell^{-1}(\beta_{(+)}-\gamma_{(+)}) \Bigl[ 
1+\delta a_0^4 +\frac{\ell}{6} \kappa^2 (\beta_{(+)}-\gamma_{(+)}) \frac{(1+\delta a_0^4)^2}{1-a_0^2}
\Bigr] h_{\mu\nu} \nonumber \\
& & +\kappa^2 \ell^{-1} (\beta_{(+)}-\gamma_{(+)})
\Biggl[
\frac{109}{56}\frac{1+\delta a_0^4}{1-a_0^2} (a_0^{-14}-1)
+\frac{3}{8}(1+\delta a_0^4)+(1+\delta)a_0^{-12}
\Biggr]T_{\mu\nu}^{(+)}
\nonumber \\
& & -\frac{3}{8}\ell^{-1} \Bigl(\frac{1}{14}(1-a_0^{-14})+\frac{1}{12}(1-a_0^2) \Bigr)
\Bigl({\cal D}_\alpha {\cal D}_\nu T^{\mu (+)}_\alpha +{\cal D}_\alpha {\cal D}^\mu T^{(+)}_{\nu\alpha}-{\cal D}^2 T^{\mu (+)}_\nu \Bigr).  
\end{eqnarray}
%
%=================================%
Then the effective gravitational equation with respect to the induced metric on the $D_+$-brane 
is given as
%===========<Equation>============%
%
\begin{eqnarray}
(1-a_0^2) {}^{(4)}G_{\mu\nu}(g_{(+)})
& = &  -\kappa^2 \ell^{-1}(\beta_{(+)}-\gamma_{(+)}) \Biggl[ 
1+\delta a_0^4 +\frac{\ell}{6}\kappa^2 (\beta_{(+)}-\gamma_{(+)})\frac{(1+\delta a_0^4)^2}{1-a_0^2} \Biggr] g_{(+)\mu\nu} 
\nonumber \\ 
& & + \frac{2}{\ell}a_0^2 \Biggl[\stac{(+)}{D_\mu} \stac{(+)}{D_\nu}d 
-g_{(+) \mu\nu} \stac{(+)}{D^2}d 
+\frac{1}{\ell} \Bigl \lbrace \stac{(+)}{D_\mu} d \stac{(+)}{D_\nu}d 
+\frac{1}{2} g_{(+)\mu\nu} (\stac{(+)}{D} d)^2 \Bigr\rbrace
\Biggr] \nonumber \\
& & +\kappa^2 \ell^{-1} (\beta_{(+)}-\gamma_{(+)})
\Biggl[
\frac{109}{56}\frac{1+\delta a_0^4}{1-a_0^2} (a_0^{-14}-1)
+\frac{3}{8}(1+\delta a_0^4)+(1+\delta)a_0^{-12}
\Biggr]T_{\mu\nu}^{(+)}
\nonumber \\
& & -\frac{3}{8}\ell^{-1} \Bigl(\frac{1}{14}(1-a_0^{-14})+\frac{1}{12}(1-a_0^2) \Bigr)
\Bigl( \stac{(+)}{D}_\alpha \stac{(+)}{D}_\nu T^{\mu (+)}_\alpha 
+ \stac{(+)}{D}_\alpha \stac{(+)}{D^\mu} T^{(+)}_{\nu\alpha}
- \stac{(+)}{D^2} T^{\mu (+)}_\nu \Bigr).  
\end{eqnarray}
%
%=================================%
Now we can see that the gauge field can be a source for the gravity on the brane.
This low energy effective theory is the four dimensional Einstein-Maxwell theory
with the cosmological constant and radion. Further, surprisingly, the effective gravitational
constant and the cosmological constant are proportional to the same factor,
$\beta_{(+)}-\gamma_{(+)}$. Noting that $\beta_{(+)}-\gamma_{(+)}$ vanishes when
$\beta_{(-)}-\gamma_{(-)}$ vanishes (see Eqs. (\ref{alphamunu}) and (\ref{betamunu})),
this means that if there is no net cosmological constant on the brane, the coupling
between the Maxwell field and the gravity also vanishes.

The equation for the radion becomes 
%===========<Equation>============%
%
\begin{eqnarray}
& & a^2_0 \Bigl(-\stac{(+)}{D^2} d +\frac{1}{\ell} (\stac{(+)}{D}d )^2  \Bigr)
-\frac{2}{3}\kappa^2 (\beta_{(+)}-\gamma_{(+)}) 
-\frac{2}{3}\kappa^2 a_0^2 (\beta_{(-)}-\gamma_{(-)}) 
-\frac{\ell}{18} \kappa^4 (\gamma_{(+)}-\beta_{(+)})^2 \Bigl( \frac{1+\delta a_0^4}{1-a_0^2}\Bigr)^2(a_0^{-2}-1)  \nonumber \\ 
& & ~~~= -\frac{\kappa^2}{2}(\beta_{(+)}-\gamma_{(+)})  (1-a_0^{-2}) \stac{(+)}{{\cal F}_{\mu\nu}}
 \stac{(+)}{{\cal F}^{\mu\nu}}(h).
\end{eqnarray}
%
%=================================%
\end{widetext}
They are main results in our paper.

\section{Summary and Discussion}

In this paper we considered a Randall-Sundrum type I model \cite{RSI} based on the IIB supergravity.
In our model the bulk spacetime and brane are described by a mimic of the IIB supergravity 
compactified on $S^5$ and the Born-Infeld plus Chern-Simons action. The brane charge is not set
equal to the brane tension (non-BPS condition). Then we derived an effective gravitational 
equation on the branes using the gradient expansion method (long wave approximation). 
By virtue of non-BPS condition, the gauge field localized on the brane 
can be a source for the gravity on the brane and the cosmological constant is induced 
at the same time. An interesting point here is that the cosmological constant is proportional
to the effective gravitational coupling to the gauge field. Then the effective theory is
the four dimensional Einstein-Maxwell theory with a cosmological constant and radion field. 
In this sense we could obtain an acceptable model for the real universe, although the contribution
from radion might be phenomenologically dangerous but would vanish if the radion could be stabilised. 

According to our result and assuming $a_0 \ll 1$, the effective gravitational coupling
$G_{(+)}^{\rm eff}$ and the net cosmological constant $\Lambda_{(+)}$ are approximately given by 
%===========<Equation>============%
%
\begin{eqnarray}
G_{(+)}^{\rm eff} \sim \frac{\kappa^2}{\ell}\frac{\beta_{(+)}-\gamma_{(+)}}{\beta_{(+)}} a_0^{-14}
\end{eqnarray}
%
%=================================%
and
%===========<Equation>============%
%
\begin{eqnarray}
\Lambda_{(+)} \sim \frac{\kappa^2}{\ell} (\beta_{(+)}-\gamma_{(+)}),
\end{eqnarray}
%
%=================================%
respectively. Using $\Lambda_{(+)}$, $G_{(+)}^{\rm eff}$ can be written as 
$G_{(+)}^{\rm eff} \sim a_0^{-14}\Lambda_{(+)} \beta_{(+)}^{-1}$. 
Assuming $ \Lambda_{(+)} \sim H_0^2 \sim (10^{-42}{\rm GeV})^2$, 
$\ell \sim (\kappa^2 \beta_{(+)})^{-1} \sim 0.1 {\rm mm}$ \cite{Exp} 
and $M_5:=\kappa^{-2/3} \sim {\rm TeV}$, 
we obtain 
%===========<Equation>============%
%
\begin{eqnarray}
a_0^{-1} \sim 10^{8.5} \Bigl( \frac{0.1{\rm mm}}{\ell}\Bigr)^{\frac{1}{14}}
\Bigl( \frac{H_0^2}{\Lambda_{(+)}} \Bigr)^{\frac{1}{14}}
\Bigl( \frac{M_5}{{\rm TeV}}\Bigr)^{\frac{3}{14}}
\end{eqnarray}
%
%=================================%
or $y_0/\ell \sim 19.5$. $H_0$ is the present Hubble constant. Hence, we could have appropriate
values for $G_{(+)}^{\rm eff}$ and $\Lambda_{(+)}$ if $y_0/\ell \sim 19.5$. 
Here we must have a fine tuning between the brane tension and charge as
$ \frac{\beta_{(+)}-\gamma_{(+)}}{\beta_{(+)}} \sim 10^{-60}(\ell /0.1{\rm mm})^2$,
which may correspond to the cosmological constant problem.
This indicates the relation of the presence of the cosmological constant and 
the gravitational coupling. In this model, the effective gravitational coupling is turned off 
if we are living on the BPS D-brane, that is, the presence of the gravitational coupling 
indicate the presence of the cosmological constant. In classical level the cosmological constant 
appears as a result of breaking BPS condition for the brane configuration. 
What we have to explain is 
the appropriate radion stabilisation mechanism so that $y_0/\ell \sim 19.5$. 

In this paper, we considered only the bosonic part for simplicity. If we want a phenomenologically 
acceptable model where the matter sector reduces to the grand unified theory at low energy,
we need to think of, for example, D3-D7 system \cite{IWT}. To investigate 
this kind of systems we must treat the brane with higher co-dimensions containing 
the fermionic parts as well as bosonic parts. This issue is left for future study.

%======================================%
%<<<<<<<<< Acknowledgements  >>>>>>>>>>>%
%======================================%
%\baselineskip25pt

\section*{Acknowledgements}

We thank Kei-ichi Maeda for his comment. 
YI and SF thank Shinya Tomizawa and Akihito Shirata for their discussions. 
The work of TS was supported by Grant-in-Aid for Scientific 
Research from Ministry of Education, Science, Sports and Culture of 
Japan(No.13135208, No.14740155 and No.14102004). 
The work of KT was supported by JSPS.


\begin{thebibliography}{22}

\bibitem{Review}
G. Gabadadze, {\it ICTP lectures on large extra dimensions}, hep-ph/0308112; 
R. Maartens, Living Rev. Rel. {\bf 7}, 7 (2004); 
P. Brax, C. van de Bruck and A. Davis, Rept. Prog. Phys. {\bf 67}, 2183(2004); 
C. Csaki, {\it TASI Lectures on extra dimensions and branes}, hep-ph/0404096. 

\bibitem{Mirage}
A. Kehagias and E. Kiritsis, JHEP {\bf 9911}, 022(1999).  

\bibitem{DBW1}
C. P. Burgess, P. Martineau, F. Quevedo and R. Rabadan, JHEP {\bf 06}, 037(2003);
C. P. Burgess, N. E. Grandi, F. Quevedo and R. Rabadan, JHEP {\bf 0401}, 067(2004);
K. Takahashi and K. Ichikawa, Phys. Rev. {\bf D69},103506(2004). 

\bibitem{DBW2}
T. Shiromizu, T. Torii and T. Uesugi, Phys. Rev. {\bf D67}, 123517(2003);
M. Sami, N. Dadhich and T. Shiromizu, Phys. Lett. {\bf B568},118(2003);
E. Elizalde, J. E. Lidsey, S. Nojiri and S. D. Odintsov, Phys. Lett. {\bf B574}, 1(2003);
T. Uesugi, T. Shiromizu, T. Torii and K. Takahashi, Phys. Rev. {\bf D69}, 043511(2004). 

\bibitem{DBW3}
S. B. Giddings, S. Kachru and J. Polchinski, Phys. Rev. {\bf D66}, 106006(2002);
O. DeWolfe and S. B. Giddings, Phys. Rev. {\bf D67}, 066008(2002). 

\bibitem{Dbrane}
S. Kachru, R. Kallosh, A. Linde, J. Maldacena, L. McAllister and S. P. Trivedi, JCAP {\bf 0310},013(2003).

\bibitem{SKOT}
T. Shiromizu, K. Koyama, S. Onda and T. Torii, Phys. Rev. {\bf D68}, 063506(2003).

\bibitem{OSKH}
S. Onda, T. Shiromizu, K. Koyama and S. Hayakawa,  Phys. Rev. {\bf D69},123503(2004). 

\bibitem{SHT}
T. Shiromizu, Y. Himemoto and K. Takahashi, Phys.Rev. {\bf D70}, 107303(2004); 
T. Shiromizu, K. Takahashi, Y. Himemoto and S. Yamamoto, Phys. Rev. {\bf D70}, 123524(2004). 

\bibitem{RSI}
L.~Randall and R.~Sundrum, Phys. Rev. Lett. {\bf 83}, 3370 (1999).

\bibitem{RSII}
L.~Randall and R.~Sundrum, Phys. Rev. Lett. {\bf 83}, 4690 (1999).

\bibitem{SKT}
T. Shiromizu, K. Koyama and T. Torii, Phys. Rev. {\bf D68}, 103513(2003).  

\bibitem{TS}
K. Takahashi and T. Shiromizu,  Phys. Rev. {\bf D70}, 103507(2004).  

\bibitem{GE}
T. Wiseman, Class. Quant. Grav. {\bf 19}, 3083(2002);
S. Kanno and J. Soda, Phys. Rev. D{\bf 66}, 043526(2002);
{\rm ibid}, 083506,(2002);
T. Shiromizu and K. Koyama, Phys. Rev. D{\bf 67}, 084022(2003);
S. Kanno and J. Soda, Gen. Rel. Grav. {\bf 36}, 689(2004). 

\bibitem{SI}
T. Shiromizu and D. Ida, Phys. Rev. {\bf D64}, 044015(2001).  

\bibitem{moduli}
J. Khoury, B. A. Ovrut, P. J. Steinhardt and N. Turok, Phys. Rev. {\bf D64}, 123522(2001);
J. Garriga, O. Pujolas and T. Tanaka, Nucl. Phys. {\bf B655}, 127(2003);
P. Brax, C. van de Bruck, A. C. Davis and C. S. Rhodes, Phys. Rev. {\bf D67}, 023512(2003);
G. A. Palma and A. C. Davis, Phys. Rev. {\bf D70}, 106003(2004); 
S. L. Webster and A. C. Davis, hep-th/0410042;
S. Kanno and J. Soda, hep-th/0410061.  

\bibitem{SMS}
T. Shiromizu, K. Maeda and M. Sasaki, Phys. Rev. {\bf D62}, 024012(2000).

\bibitem{Exp}
C. D. Hoyle, D. J. Kapner, B. R. Heckel, E. G. Adelberger, J. H. Gundlach, 
U. Schimidt and H. E. Swanson, Phys. Rev. {\bf D65}, 044023(2002). 

\bibitem{IWT}
Y. Imamura, T. Watari and T. Yanagida, Phys. Rev. {\bf D64}, 065023(2001);
T. Watari and T. Yanagida, Phys. Rev. {\bf D70}, 036009(2004). 

\end{thebibliography}
\end{document}